\documentclass[11pt]{article}
\usepackage{moriond,epsfig}
\usepackage{wrapfig}

\def\be{\begin{equation}}
\def\ee{\end{equation}}
\def\bea{\begin{eqnarray}}
\def\eea{\end{eqnarray}}

\def\ipb{\,pb\ensuremath{^{-1}}}
\def\ifb{\,fb\ensuremath{^{-1}}}

\def\eminp{$e^{-}p$}
\def\eplusp{$e^{+}p$}
\def\eplusminp{$e^{\pm}p$}

\def\qsq{\ensuremath{Q^{2}}}

\def\ptx{$P_{T}^{X}$}
\def\pole{$P_{e}$}
\def\fmin{$F_{-}$}
\def\fzero{$F_{0}$}
\def\chisq{$\chi^{2}$}

\def\vu{$v_{u}$}
\def\au{$a_{u}$}

\def\w{$W$}
\def\wb{\w\ boson}
\def\z{$Z$}
\def\zb{\z\ boson}
\def\heraone{HERA I}
\def\heratwo{HERA II}
\def\heraoneplustwo{HERA I+II}

\newcommand{\cosths}{\ensuremath{\cos\,\theta^{*}}}

\newcommand{\mis}[1]{{\bf /}\!\!\!\!#1}
\newcommand{\ptmiss}{\ensuremath{\mis{P}_{T}}}

\newcommand{\isolep}{\ensuremath{\ell+\ptmiss}}

\newcommand{\dSigmaSMdQ}{\ensuremath{\textrm{d}\sigma_{SM}/\textrm{d}\qsq}}
\newcommand{\quarkradius}{\ensuremath{R_{q}}}

\newcommand{\honequarklimit}{ \ensuremath{0.74\cdot 10^{-18}} }
\newcommand{\zeusquarklimit}{ \ensuremath{0.62\cdot 10^{-18}} }
\newcommand{\xsecw}{\ensuremath{\sigma_{W} = 1.2 \pm 0.3 \, \left(\textrm{stat}\right) \pm 0.2 \, \left(\textrm{sys}\right) \, \textrm{pb}}}
\newcommand{\xsecwsm}{\ensuremath{1.3\pm 0.2\,\textrm{pb}}}

\begin{document}
\vspace*{4cm}
\title{ELECTROWEAK MEASUREMENTS FROM HERA}

\author{Y. R. de Boer\\
$\phantom{niks}$\\
On behalf of the H1 and ZEUS collaborations}

\address{$\phantom{niks}$\\
DESY\\
Notkestrasse 85\\
22607 Hamburg, Germany}

\maketitle\abstracts{
New preliminary electroweak results from the HERA lepton-proton collider
experiments H1 and ZEUS are presented.
These include new high \qsq\ neutral current cross section measurements,
limits on a possible quark radius
in the search for contact interactions as well as on
quark-\z\ coupling parameters, extracted
in combined electroweak and QCD fits.
Furthermore, new 
charged current cross section measurements
as a function of the lepton-beam polarisation are presented,
as well as charged current measurement results,
using the combined \mbox{\heraone} data from H1 and ZEUS.
Finally, measurements of the single \wb\ production cross section
and the \wb\ polarisation fractions are presented.
}

\section{Introduction}
The lepton-proton collider HERA~\cite{Schmueser:1984hd} has facilitated
measurements of electroweak (EW) interactions between quarks
and leptons in deep inelastic scattering (DIS)
at a centre of mass energie up to 320\,GeV and four momentum of the exchanged boson squared
(\qsq) up to 40000\,GeV$^{2}$.
The data taking took place 
in the years 1994-2000 (\heraone) and 2003-2007 (\heratwo).
Two collider-mode detectors, H1~\cite{Abt:1996hi} and
ZEUS~\cite{zeus.detector},
have each collected approximately $0.5$\ifb\ of data, divided in electron-proton (\eminp) and
positron-proton (\eplusp) data.
Previously obtained HERA results have led to a significantly improved understanding
of the proton substructure~\cite{Adloff:2000qk,Chekanov:2002pv}
allowing for measurements of important EW
parameters~\cite{Aktas:2005iv,Adloff:1999ah,Breitweg:1999aa,Chekanov:2006da}
as well as searches~\cite{Adloff:2003jm,Breitweg:1999ssa} for contact interactions.
In these proceedings, updates are presented to these analyses
along with new results from H1 regarding the single production of \wb s at HERA.


\begin{figure}[t]

\begin{center}
  \begin{minipage}[t]{.49\linewidth}
    \par \vspace{0mm}
    \includegraphics[width=\linewidth]{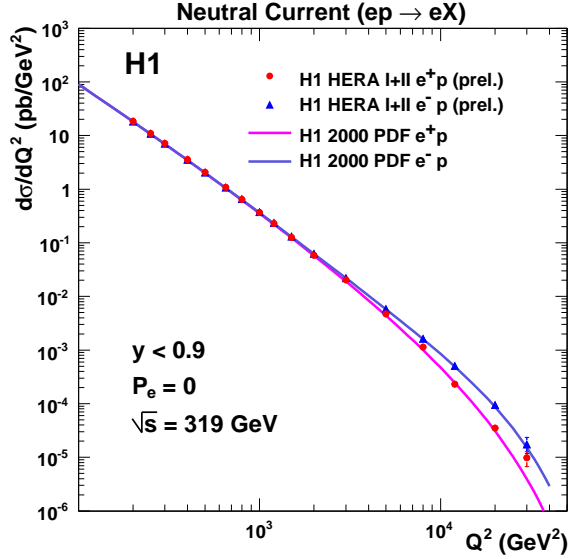}
  \end{minipage}
\end{center}

\vspace{-10mm}
\caption{NC single differential cross sections as a function of \qsq, measured by H1 using \eplusp\ (points) and \eminp\ (triangles) DIS data
  from the full \heraoneplustwo\ data set, compared to the SM QCD expectations
  (solid lines).
  The interference between the photon and \zb, which is different in \eplusp\ and \eminp\ collisions,
  becomes visible at \qsq\ of the order $10^{4}$\,GeV$^{2}$.
}\label{fig:nc.vs.q2.xsec}

\end{figure}

\section{Neutral Current Cross Section Measurement}
New H1 Neutral Current (NC) single differential cross section measurements~\cite{h1.nc.quark.radius}
as a function of \qsq, are presented in Figure~\ref{fig:nc.vs.q2.xsec}.
The analysis includes previously published~\cite{Adloff:1999ah,Adloff:2000qj,Adloff:2003uh}
\heraone\ and preliminary~\cite{aktas.ichep06} \heratwo\ data,
corresponding to an integrated luminosity of
270\ipb\ of \eplusp\ data and 165\ipb\ of \eminp\ data.
The measurement precision is better than $10\%$ for \qsq\ up to
20000\,GeV$^{2}$.
The data agree well with SM QCD expectations, which are
based on parton distribution functions obtained
using high energy \heraone\ data.~\cite{Adloff:2000qk}

\section{Derivation of Limits on the Quark Radius in Contact Interactions}
In the search for contact interactions, both H1 and ZEUS 
derive limits on a possible quark radius
using high \qsq\ NC events.~\cite{h1.nc.quark.radius,zeus.quark.radius}
A form factor $f_{q}$, as a function of \qsq\ and a hypothetical quark radius 
\quarkradius, is defined as
$f_{q}\left(\qsq,\quarkradius\right) \equiv 1 - \frac{1}{6}\left<\quarkradius^{2}\right>\qsq$.
This leads to an altered single differential cross section
$\textrm{d}\sigma/\textrm{d}\qsq =
f_{q}\left(\qsq,\quarkradius\right)\,$\dSigmaSMdQ\ for contact interactions,
which can be fit to the data, as is shown in Figure~\ref{fig:quark.radius}.
The 95\% Confidence Level (CL) limits on the quark radius are determined from the fit to be
\honequarklimit m and \zeusquarklimit m, by H1 and ZEUS, respectively.
These limits confine a hypothetical quark radius to be less than
or equal to the resolution power of HERA.
\begin{figure}[h]
  \begin{minipage}[t]{.49\linewidth}
    \par \vspace{0mm}
    \includegraphics[width=\linewidth]{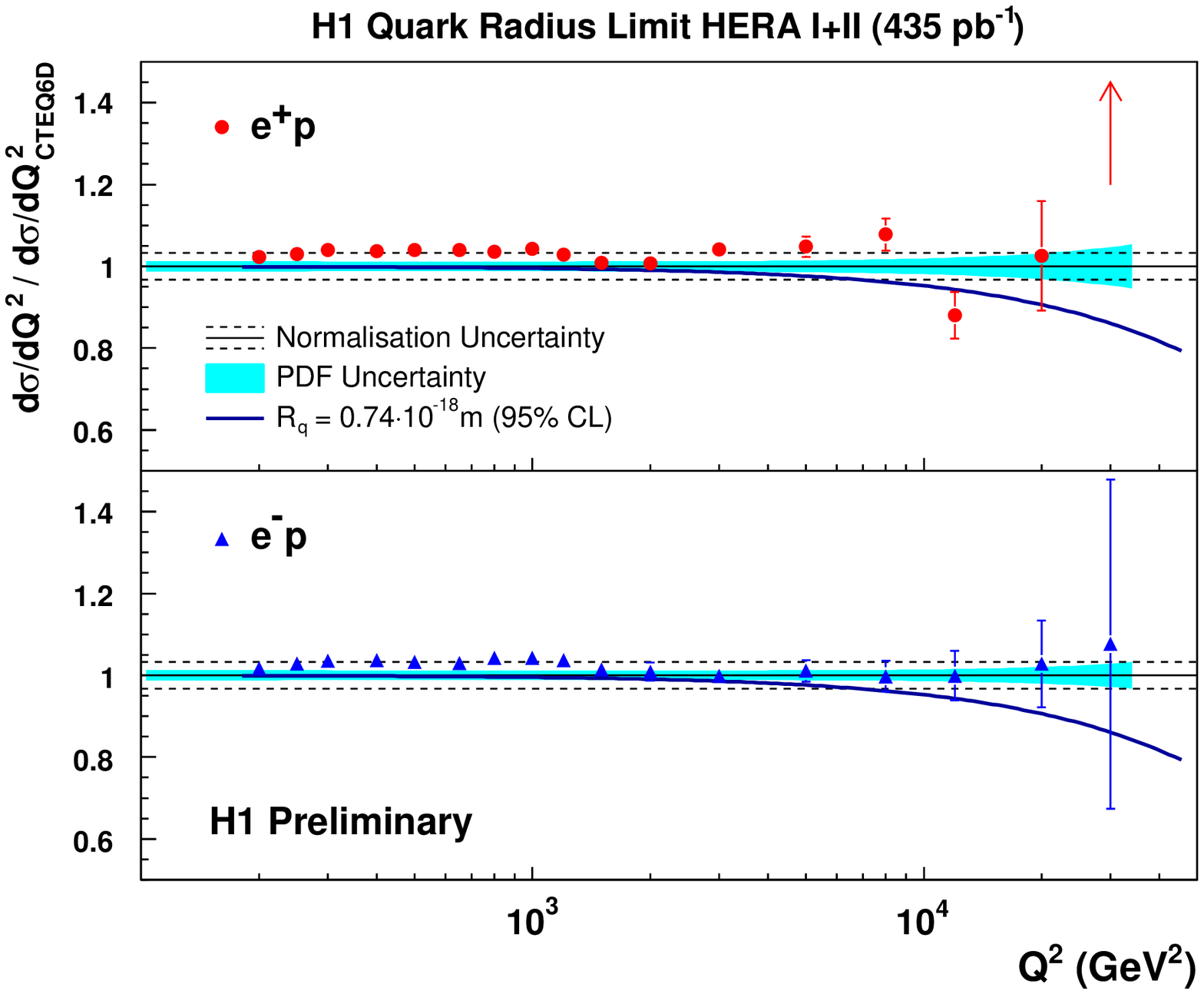}
  \end{minipage}
  \hfill
  \begin{minipage}[t]{.49\linewidth}
    \par \vspace{0mm}
    \includegraphics[width=\linewidth]{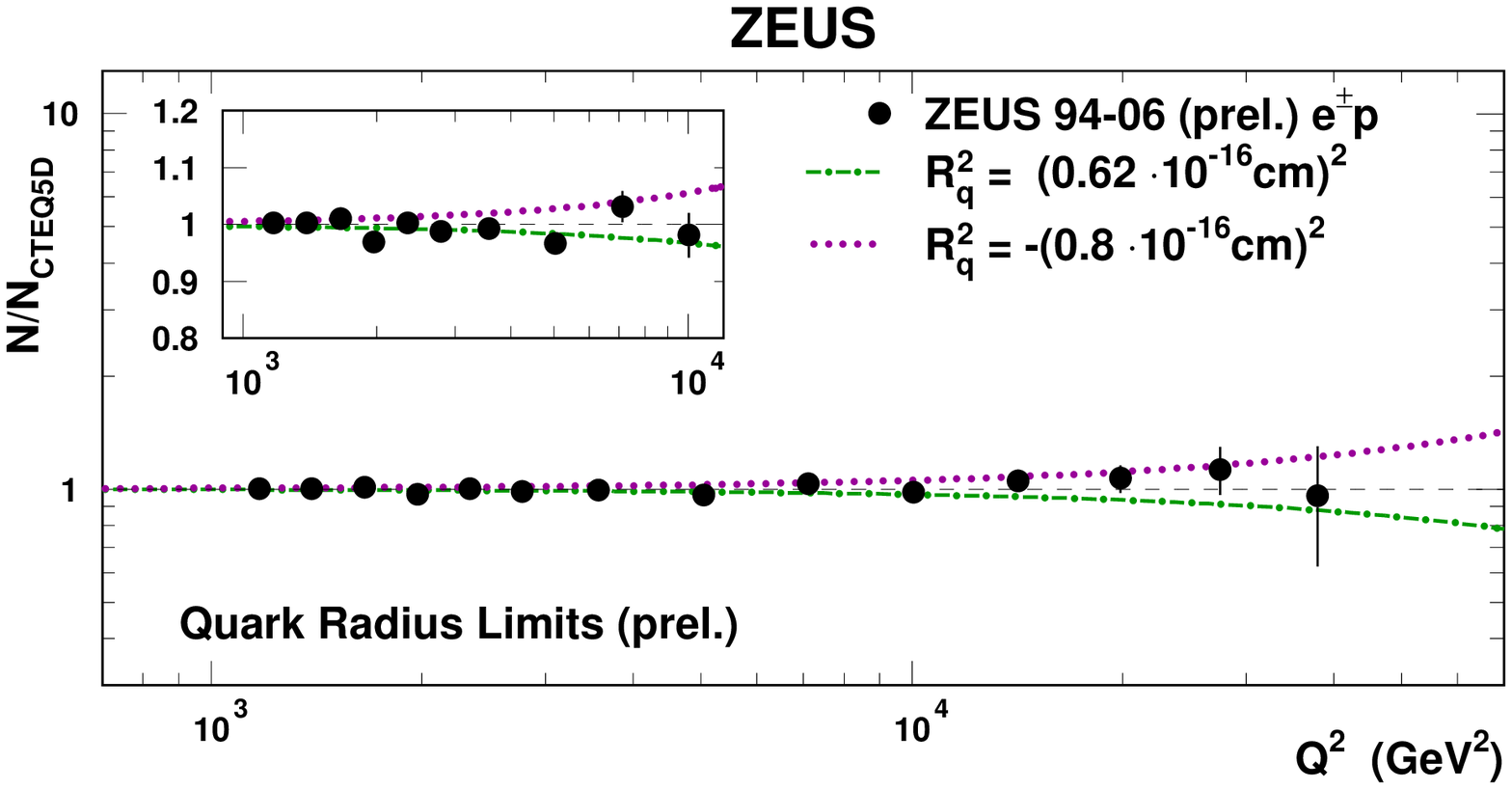}
  \end{minipage} 
\vspace{-3mm}
\caption{NC single differential cross sections as a function of \qsq\ normalised to the SM expectation \dSigmaSMdQ.
The lines represent corrections 
due to the hypothetical quark radius \quarkradius\ at its 95\% CL limit for
H1 \eplusp\ data (top left), H1 \eminp\ data (bottom left), and for ZEUS \eplusminp\ (right). }
\label{fig:quark.radius}
\end{figure}

\section{Combination of H1 and ZEUS Charged Current \heraone\ Data}
The H1 and ZEUS collaborations are combining their data
to improve the precision of DIS measurements.~\cite{h1.zeus.combo}
New preliminary results, using combined \heraone\ data of H1 and ZEUS, 
are shown in Figure~\ref{CC.eplusminp}, depicting
CC reduced cross sections in bins of \qsq.
A good agreement is observed between the combined data~\cite{h1.zeus.cc.xsecs}
and the QCD fits of each experiment~\cite{Adloff:2000qk,Chekanov:2005nn} to their own data.
The \eplusp\ data correspond
to a total integrated luminosity of approximately $200$\ipb\ and
have a typical precision of $8\%$. 
The statistical gain in precision is most significant in the 
statistically limited \eminp\ data set,
where the combined data ($30$\ipb) leads to an increase of the precision
to about 20\%.
The precision is expected to increase further with the future
inclusion of the \heratwo\ data.

\begin{figure}[t]

  \begin{minipage}[t]{.49\linewidth}
    \par \vspace{0mm}
    \includegraphics[width=\linewidth]{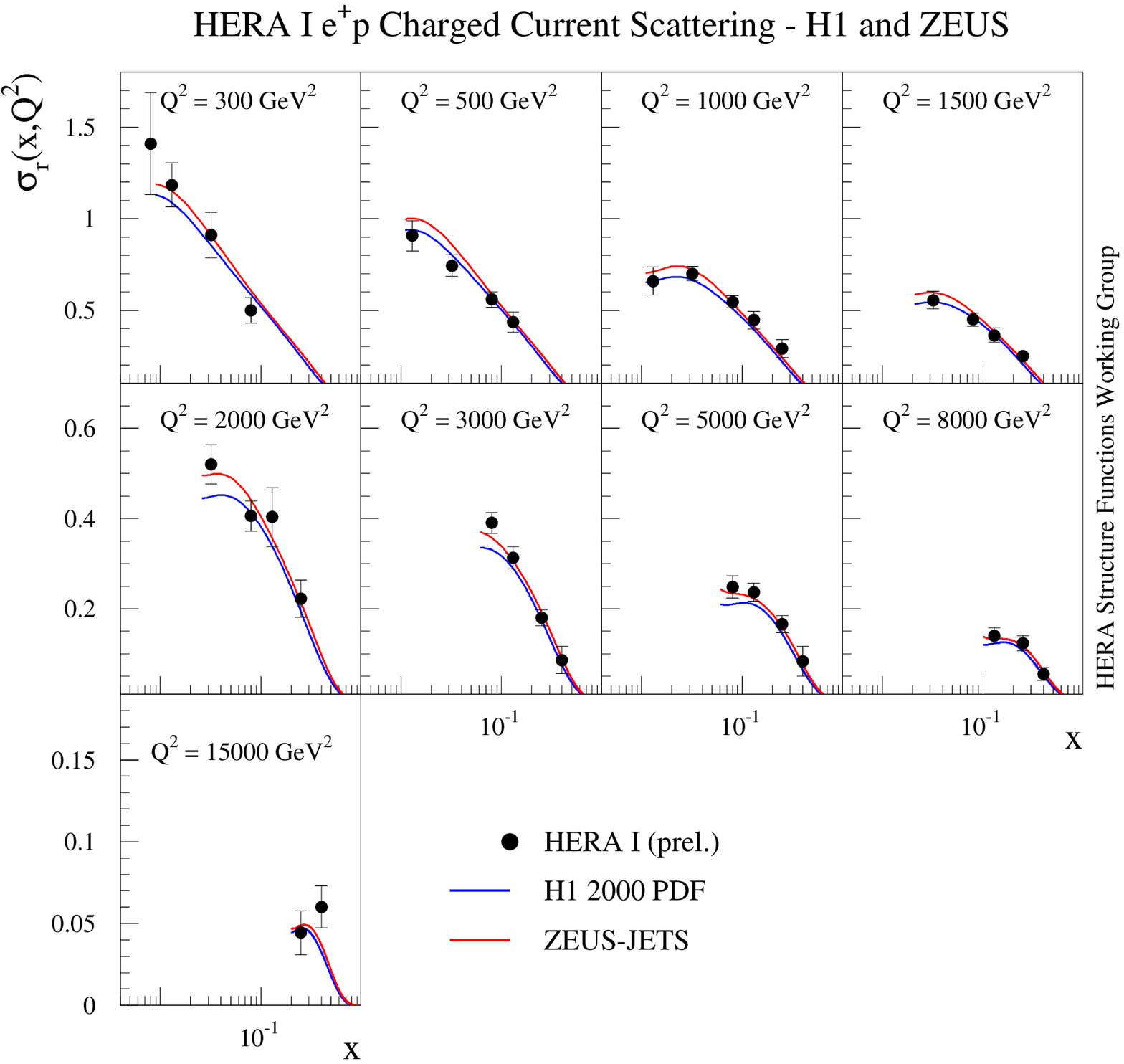}
  \end{minipage}
  \hfill
  \begin{minipage}[t]{.49\linewidth}
    \par \vspace{0mm}
    \includegraphics[width=\linewidth]{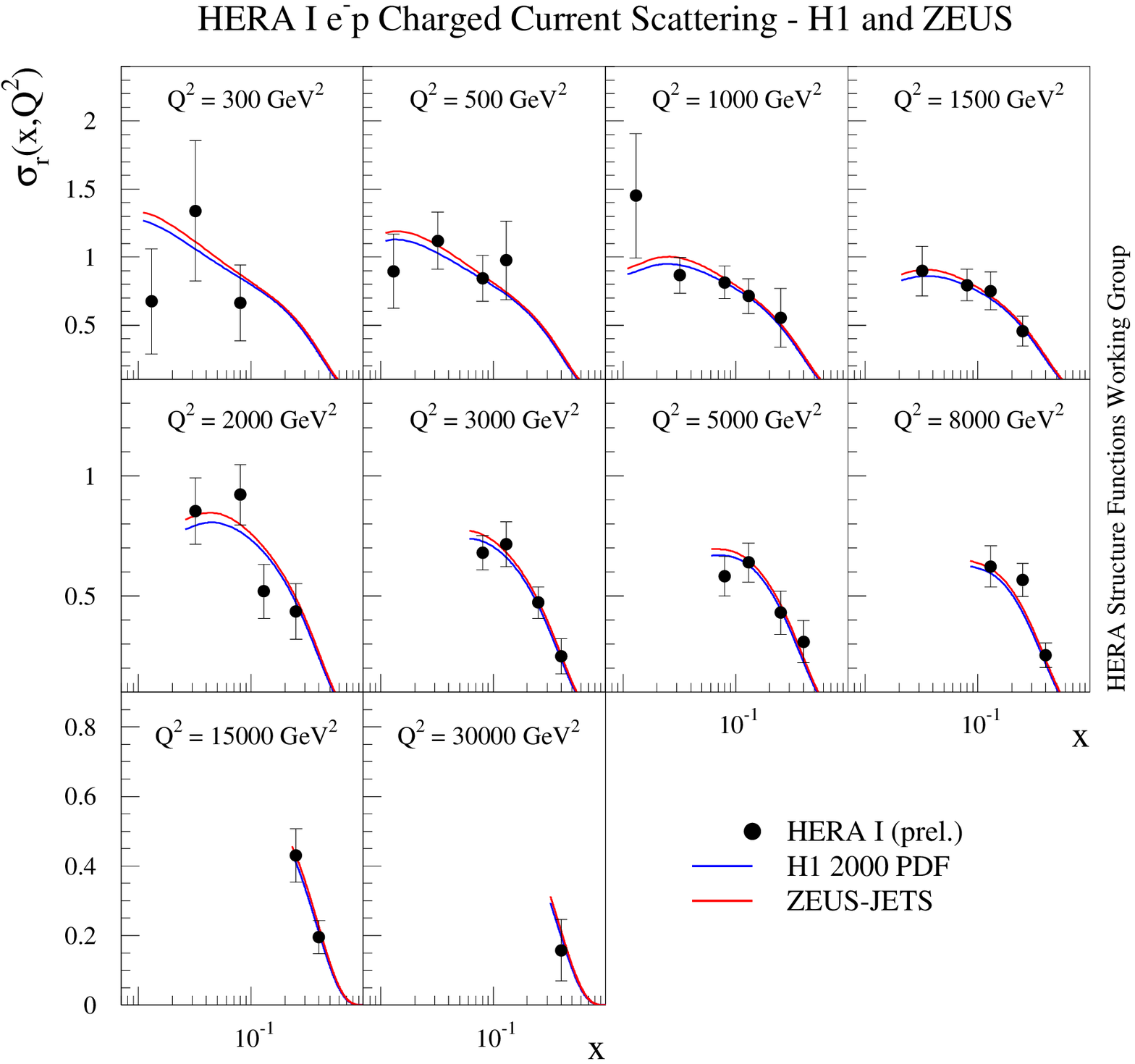}
  \end{minipage} 

\vspace{-6mm}
\caption{Reduced CC cross sections (points), using the combined \heraone\ data of H1 and ZEUS,
  in bins of \qsq\ for \eminp\ (left)
  and \eplusp\ (right) data. The curves are NLO QCD fits, performed by H1 and ZEUS to their own data.}
\label{CC.eplusminp}

\end{figure}

\section{Charged Current Cross Section Measurement using Polarised Lepton-Beams}
During the \heratwo\ running period, 
longitudinally polarised lepton-beams were used.
The polarisation is defined as
\pole$\equiv\left(N_{R}-N_{L}\right)/\left(N_{R}+N_{L}\right)$,
with $N_{R}\left(N_{L}\right)$ the number of right (left)
handed leptons in the beam.
The SM predicts a linear scaling of the CC cross section
with the beam polarisation, 
due to the absence of a right handed neutrino.
New ZEUS results,~\cite{zeus.cc.pol} using data from the years 2006-2007, 
are shown in Figure~\ref{CC.pol}, together with previously obtained HERA
measurements.~\cite{Aktas:2005ju,Chekanov:2006da}
A good agreement between the  measurements in the different data sets and the SM expectations is observed.

\begin{figure}[tb]

  \begin{minipage}[t]{.49\linewidth}
    \par \vspace{1.5mm}
    \includegraphics[width=\linewidth]{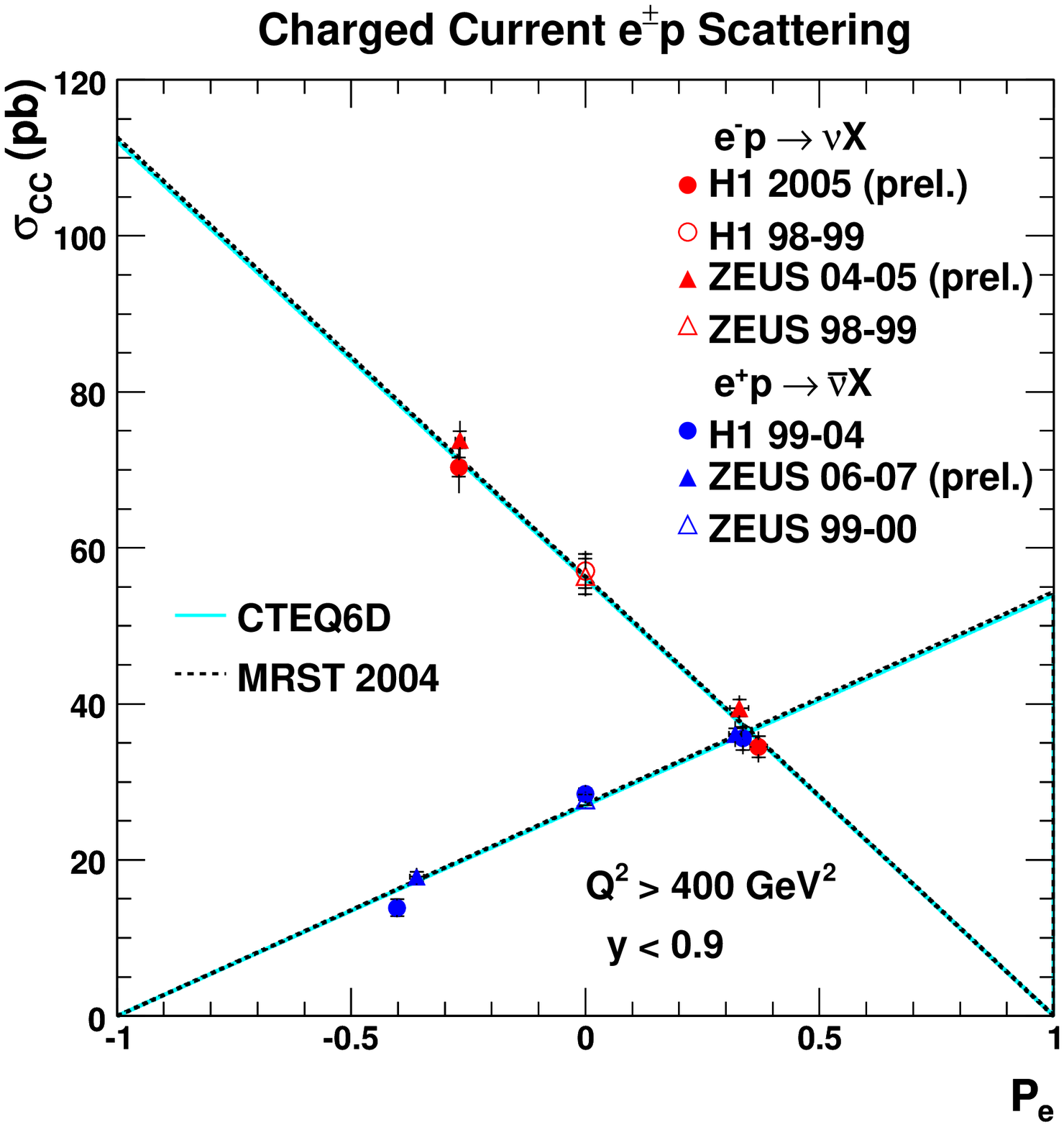}
    \caption{The polarisation dependence of the
      charged current cross section
      on the lepton-beam polarisation \pole. The curves are the SM QCD predictions
      using two different PDF fits.}
    \label{CC.pol}
  \end{minipage}
  \hfill
  \begin{minipage}[t]{.49\linewidth}
    \par \vspace{1.5mm}
    \includegraphics[width=\linewidth]{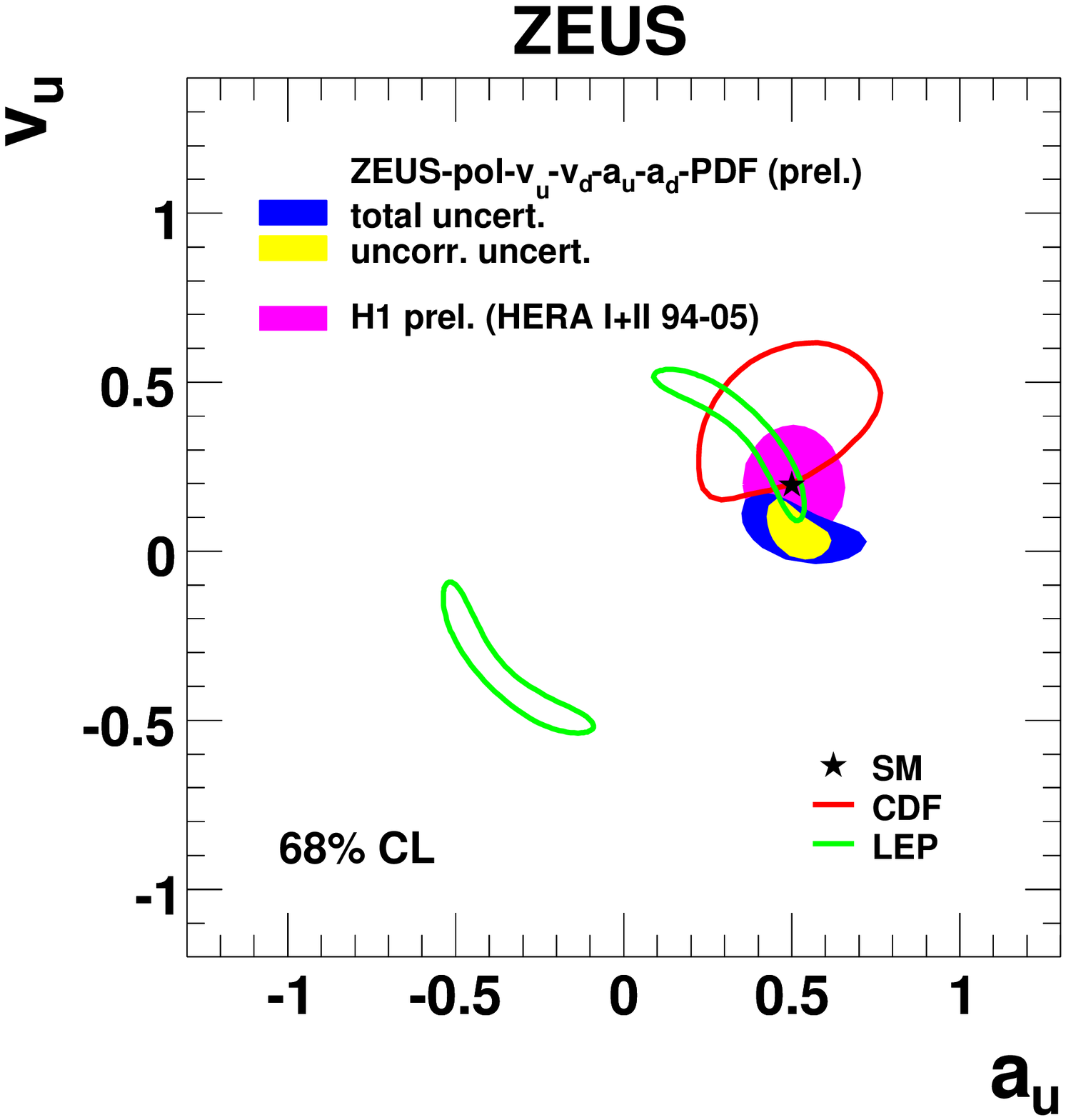}
    \caption{Limits at 68\% CL on the vector and axial-vector couplings of the \zb\ to the up quark, \au\ and \vu\
      respectively,
      shown for H1 and ZEUS in comparision with other experiments and the SM prediction.}
    \label{auvu}
  \end{minipage} \\


\end{figure}

\section{Measurement of the quark-\z\ coupling}
The cross section of NC DIS events composes of photon ($\gamma$) and
\z\ exchange diagrams. Due to the heavy \zb\ propagator, the contribution from
pure \z\ exchange is suppressed. The contribution from $\gamma Z$ interference,
however, is still sensitive to the vector and axial-vector couplings
of the \zb\ to the quark.
Limits at 68\% CL on these couplings are extracted,
using combined EW and QCD fits
where the coupling parameters pertaining to both the up and down quark are left
free in the fit.~\cite{h1.combined.fits,zeus.combined.fits}
In particular, the limits concerning the couplings to the up quark (\vu\ and \au)
profit from including the polarised \heratwo\ data.
The results are shown in Figure~\ref{auvu}, in comparison with results obtained by CDF
and LEP.~\cite{:2004qh,Acosta:2004wq}
The HERA experiments provide a better measurement than the
Tevatron and resolve the ambiguity in the LEP results.

\begin{figure}[tb]

  \begin{minipage}[t]{.49\linewidth}
    \par \vspace{0mm}
    \includegraphics[width=\linewidth]{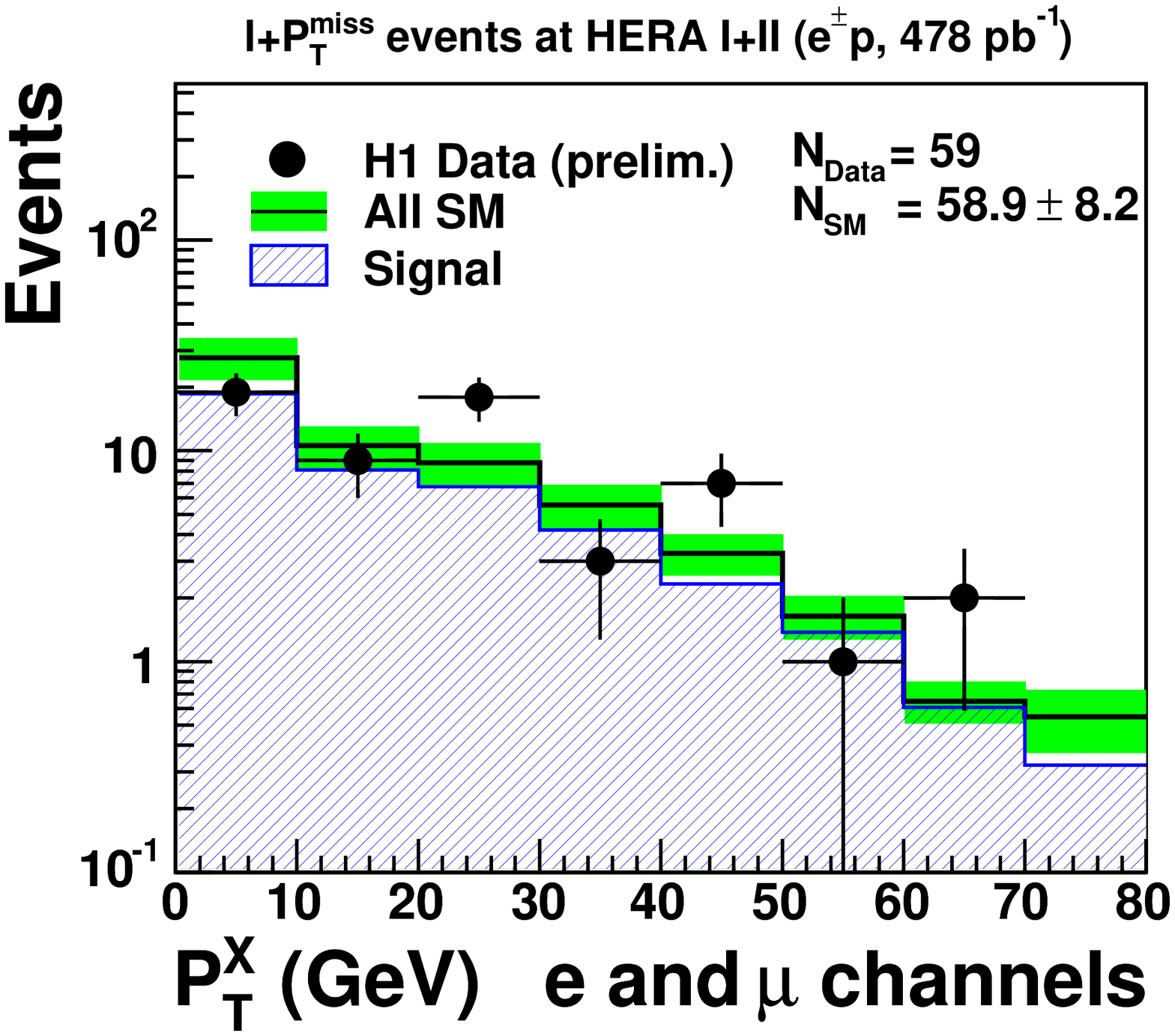}
  \end{minipage}
  \hfill
  \begin{minipage}[t]{.49\linewidth}
    \par \vspace{1mm}
    \includegraphics[width=\linewidth]{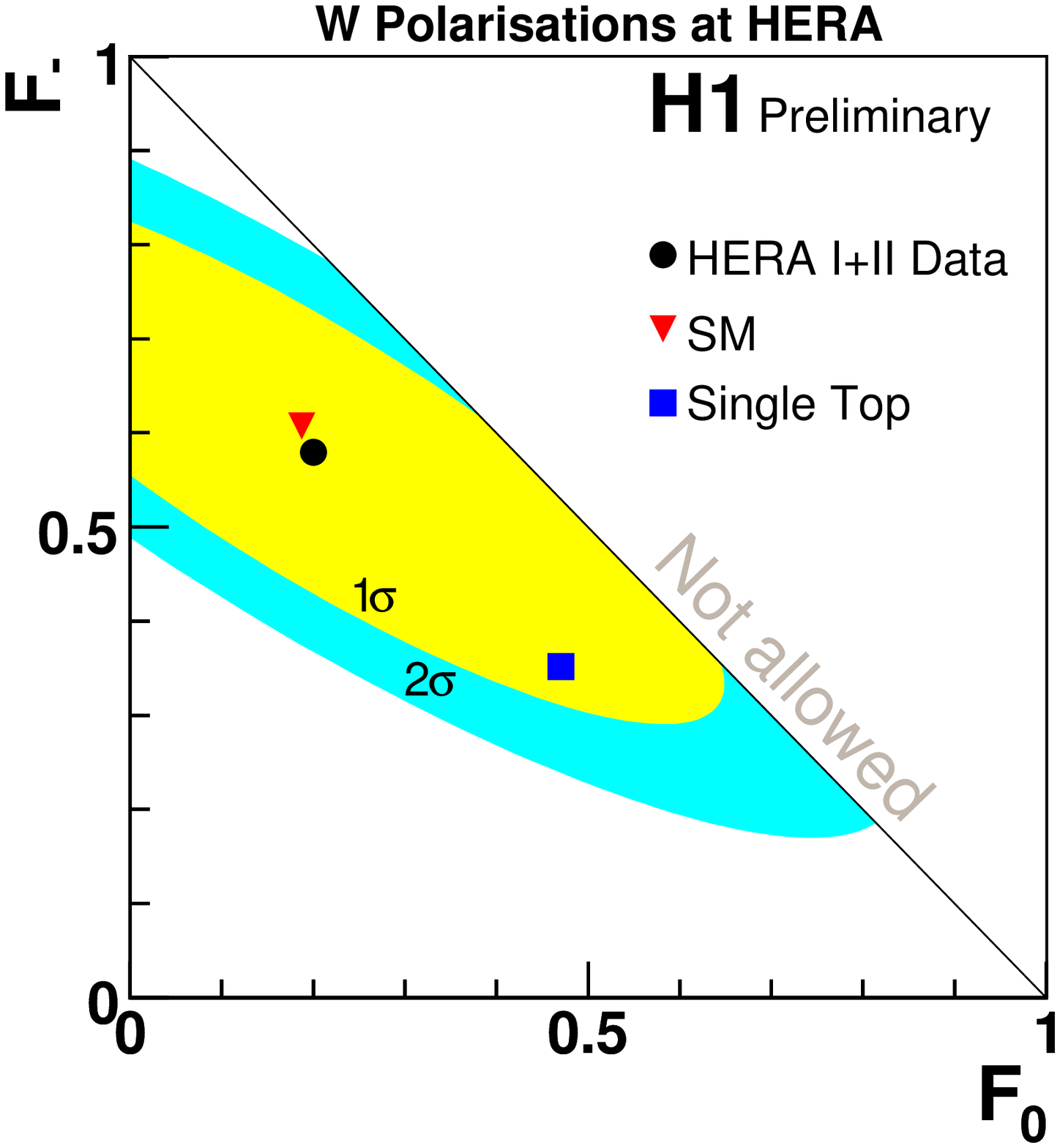}
  \end{minipage} 
  \vspace{-7mm}
  \caption{Left: The hadronic transverse momentum (\ptx) distribution in the electron
    and muon channels combined. The data (points) are compared
    to the SM expectation (open histogram). The main contribution from single \w\
    production is also shown (hatched histogram).
    Right: Measured values for the polarisation fractions \fmin\ and \fzero\ with the
    1 and 2$\sigma$ CL contours. }
  \label{fig:H1prelim-07-063.fig7.isolep.yield.vs.ptx.eps}
\end{figure}

\section{Measurement of Single \w\ Boson Production}\label{section:wxsec}
Single \wb\ production in the SM is a rare process at HERA with a cross section of order
1\,pb.~\cite{Baur:1991pp}
In the case of leptonic \wb\ decay, for which the branching ratio is about 30\%,
the event gives rise to a characteristic `\isolep' detector signature, consisting
of an energetic isolated electron or muon ($\ell$)
and large missing transverse momentum (\ptmiss).
The full \heraoneplustwo\ high energy data sample,
collected with the H1 detector in the years 1994-2007 and corresponding to an integrated 
luminosity of 478\ipb, is analysed and 
$59$ \isolep\ events are selected~\cite{isolep.prel} compared to a
SM expectation of $58.9\pm 8.2$.
This yield is presented in Figure~\ref{fig:H1prelim-07-063.fig7.isolep.yield.vs.ptx.eps} (left)
as a function of the tranverse momentum of the hadronic system (\ptx).
Notwithstanding an excess of the data over the MC
prediction in the small region of phase space where \ptx$>25$\,GeV,
a good over-all agreement with the SM is observed.
The single \wb\ production cross section is determined~\cite{w.xsec.and.pol.prelim} to be
\xsecw,
which is in good agreement with the (NLO) SM expectation of \xsecwsm.
The quoted errors on the measured cross section include
theoretical and experimental uncertainties.  

\section{Measurement of the \w\ Boson Polarisation Fractions}
\label{sec:wpolfrac}
The measurement of the \wb\ polarisation fractions is based
on the \isolep\ data sample discussed in Section~\ref{section:wxsec}
and makes use of the
\cosths\, distributions in the decay $W\rightarrow e/\mu+\nu$.
$\theta^{*}$ is defined as the angle between the \wb\ momentum in
the lab frame and that of the charged decay lepton in the \wb\ rest
frame.  For the left handed polarisation fraction $F_{-}$, the
longitudinal fraction $F_{0}$ and the right handed fraction $F_{+}
\equiv 1 - F_{-} - F_{0}$, the \cosths\, distributions for $W^{+}$
bosons are given~\cite{Hagiwara:1986vm} by
\begin{equation}
\frac{d\sigma_{W}}{d\cos\theta^{*}} 
\propto \left( 1 - F_{-} - F_{0} \right) \cdot \frac{3}{8} \left( 1 + \cos\theta^{*}\right)^{2}
+       F_{0} \cdot \frac{3}{4} \left( 1 - \cos^{2}\theta^{*}\right)
+       F_{-} \cdot \frac{3}{8} \left( 1 - \cos\theta^{*}\right)^{2}.  \label{eqn:polmodel}
\end{equation}
For $W^{-}$ bosons, the \cosths\, distributions have opposite values.
To allow the combination of both channels, \cosths\, is multiplied with
the sign of the lepton charge $q_{\ell} = \pm 1$. Therefore,
from the \isolep\ data sample, only events
for which a reliable measurement of the charge of the isolated lepton exists
are used. The reconstruction of the \wb\ rest frame is performed and the \wb\ differential
cross section as a function of the decay angle $\theta^{*}$ is derived
and fit to the model defined in Equation~\ref{eqn:polmodel}.
In the fit, the optimal values for $F_{-}$
and $F_{0}$ are simultaneously extracted using a \chisq\ minimisation method.
The result is shown in
Figure~\ref{fig:H1prelim-07-063.fig7.isolep.yield.vs.ptx.eps} (right) and
found to be in good agreement with the
SM.   $F_{-}$ and $F_{0}$ are also
extracted in fits where one parameter is fixed to its SM value. 
No deviations from the SM are observed and the values are determined to be:
\begin{eqnarray}
F_{-}  &=& 0.58  \pm \,  0.15 \, \textrm{(stat)} \pm 0.12  \,\textrm{(sys)}\quad\textrm{SM:}\quad 0.61 \pm 0.01 \, (\textrm{stat}),\nonumber \\
F_{0}  &=& 0.15 \pm \,  0.21\, \textrm{(stat)} \pm 0.09  \,\textrm{(sys)}\quad\textrm{SM:}\quad 0.19 \pm 0.01 \, (\textrm{stat}).\nonumber
\end{eqnarray}

\section{Summary}
Preliminary H1 results of NC cross section measurements at high \qsq\ have been presented,
using the complete \heraoneplustwo\ high energy data set. A good agreement with the QCD SM expectations
is observed. In the search for contact interactions,
H1 and ZEUS derive strong upper limits
on a possible quark radius 
of respectively \mbox{\honequarklimit m} and \mbox{\zeusquarklimit m} at 95\% CL.
In addition, two dimensional limits at 68\% CL on the vector and axial vector couplings
of the \zb\ to the up quark were shown,
using combined EW and QCD fits.
New measurements
of the charged current cross section as a function of the lepton-beam polarisation,
using data taken in the years 2006-2007, have been presented.
A good agreement is observed with previous measurements and with the SM,
which forbids right handed charged currents.
Recently derived CC cross section measurements are presented, using
the combined \heraone\ data of both experiments. The combination of the data has
led to significant improvements in the statistical precision.
The cross section measurements using the combined data are in good agreement with
the previously established QCD fits of H1 and ZEUS to their own data.
A single \wb\ production cross section measurement is performed by H1 using the full \heraoneplustwo\ data
and found to be \xsecw, which is in good agreement with the (NLO) SM expectation of \xsecwsm. 
Finally, the \wb\ polarisation fractions are measured and found to be 
in good agreement with the SM.

\section*{References}


\end{document}
